\documentclass[aps,pra,reprint,superscriptaddress,floatfix,longbibliography]{revtex4-2}
\usepackage[T1]{fontenc}
\usepackage{lmodern}
\usepackage[utf8]{inputenc}
\usepackage{amsmath,amssymb,bm}
\usepackage{graphicx}
\usepackage{xcolor}
\usepackage{hyperref}
\definecolor{ArticleLinkBlue}{HTML}{2E3092}
\graphicspath{{../figures/}}
\hypersetup{colorlinks=true,linkcolor=ArticleLinkBlue,citecolor=ArticleLinkBlue,urlcolor=ArticleLinkBlue,
  pdftitle={Dephasing-Powered Autonomous Quantum Engine with Time-Crystalline Work Cycles},
  pdfauthor={Zhen-huan Yang and Dan-Bo Zhang}}
\begin{document}
\title{Dephasing-Powered Autonomous Quantum Engine with Time-Crystalline Work Cycles}
\author{Zhen-huan Yang}
\affiliation{Key Laboratory of Atomic and Subatomic Structure and Quantum Control (Ministry of Education), Guangdong Basic Research Center of Excellence for Structure and Fundamental Interactions of Matter, and School of Physics, South China Normal University, Guangzhou 510006, China}
\author{Dan-Bo Zhang}
\email{dbzhang@m.scnu.edu.cn}
\affiliation{Key Laboratory of Atomic and Subatomic Structure and Quantum Control (Ministry of Education), Guangdong Basic Research Center of Excellence for Structure and Fundamental Interactions of Matter, and School of Physics, South China Normal University, Guangzhou 510006, China}
\affiliation{Guangdong Provincial Key Laboratory of Quantum Engineering and Quantum Materials, Guangdong-Hong Kong Joint Laboratory of Quantum Matter, and Frontier Research Institute for Physics, South China Normal University, Guangzhou 510006, China}
\date{November 23, 2026}

\begin{abstract}
Autonomous quantum engines seek to convert energy supplied by environments into useful work, yet dephasing is usually regarded as detrimental noise. Here we show that local dephasing can instead power a quantum engine: by disrupting ferromagnetic correlations in an interacting spin-$1/2$ ensemble, it injects energy that can be transferred through an LC resonator to a passive load. Remarkably, without any external periodic driving, the coupled system can self-organize into a time-crystalline work cycle that delivers sustained, reproducible power. Within homogeneous mean-field dynamics, we identify two distinct routes to such operation. At weak spin--resonator coupling, the dissipative time crystal of the unloaded working medium survives loading and drives the receiver. At stronger coupling, resonator backaction can disrupt cyclic operation, whereas appropriate load damping restores a stable, self-consistent periodic cycle of the composite system. Finite-size spin-diffusion simulations further reveal rapidly increasing uninterrupted operating times with different size dependences in the two regimes. Our results demonstrate how an autonomous quantum engine can self-organize its collective dynamics through interactions and load feedback to harness environmental noise for sustained cyclic work output.
\end{abstract}
\maketitle

\section{Introduction}
\label{sec:introduction}

Quantum engines link microscopic quantum dynamics with the conversion of external resources or reservoir fluctuations into useful work~\cite{Bhattacharjee2021,Allahverdyan2004,CampaioliRMP2024}. Autonomous quantum engines operate without externally imposed time-dependent control protocols, allowing energy conversion to emerge from the system's intrinsic dynamics~\cite{Carollo2020}. Sustaining cyclic output therefore naturally calls for a working medium that can self-generate collective motion and maintain it under load. Dissipation plays dual, competing roles in this process: environmental couplings can supply energy and relaxation, while decoherence can suppress the collective oscillations responsible for coherent power delivery. Understanding how these mechanisms cooperate to sustain a work-producing cycle is central to autonomous quantum energy conversion.

Continuous time crystals demonstrate how many-body interactions and dissipation can jointly sustain persistent collective oscillations under time-independent conditions~\cite{Iemini2018,Chan2015,Buca2019,Kessler2019,Kongkhambut2022,Greilich2024,Wu2024,Russo2025,Wang2025,Yang2025}. In a ferromagnetically interacting spin ensemble, local dephasing plays an especially direct energetic role: it erodes transverse ferromagnetic correlations, thereby raising the interaction energy, while its interplay with reset and collective dynamics supports persistent oscillatory motion~\cite{YangZhang2026}. This mechanism suggests harnessing environmental noise to power a repeating work cycle. However, coupling an oscillating working medium to a dynamical receiver inevitably modifies the collective motion through which energy is transferred~\cite{Carollo2020,Paulino2023,CarolloThermo2024}. This raises a fundamental question for the composite engine: can interactions and load feedback organize energy injected by time-independent dephasing noise into sustained cyclic work? Addressing this question requires treating the energy supply, collective spin dynamics, and load response self-consistently.

To address this, we investigate an interacting spin ensemble coupled to an LC resonator and a passive load with all external parameters held fixed. Within homogeneous mean-field dynamics in the thermodynamic limit, we establish the full energy balance and identify a closed work loop generated by the joint spin--resonator motion. At weak coupling, the spin limit cycle of the unloaded working medium survives loading and drives the receiver. At stronger coupling, resonator backaction can suppress accessible cyclic operation, whereas suitable load damping restores a stable composite limit cycle. Mapping the coupling strength and load damping connects these dynamical regimes to output power and energy-extraction efficiency, spanning from weak backaction to nonlinear feedback. Consequently, the load actively helps organize the collective motion through which it extracts energy.

The remainder of this article is organized as follows. Section~\ref{sec:model} introduces the microscopic model, mean-field dynamics, and energy balance. Section~\ref{sec:load_stabilization} analyzes cyclic operation, load-induced stabilization, and output waveforms. Section~\ref{sec:efficiency} examines energy-extraction efficiency and dynamical reliability across parameter space, followed by a concluding summary.

\section{Dephasing-powered spin engine}
\label{sec:model}
\subsection{Working medium and loaded dynamics}
\label{subsec:working_medium}

\begin{figure}[!htbp]
\centering\includegraphics[width=\columnwidth]{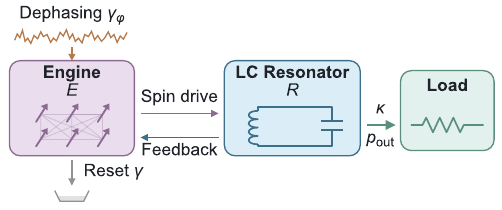}
\caption{Dephasing-powered autonomous engine.
In cyclic operation, the interacting spin ensemble $E$ forms a continuous time crystal.
Macroscopic oscillations of its longitudinal magnetization $m_z$ drive the LC resonator $R$.
The resonator displacement $u$ produces the feedback field $2gu$ on $E$, and the passive load receives power $p_{\mathrm{out}}$.}
\label{fig:engine_scheme}
\end{figure}

We consider an engine $E$ comprising $N$ spin-$1/2$ particles coupled to a single-mode LC resonator $R$ and an external passive load, as illustrated in Fig.~\ref{fig:engine_scheme}. Prior to connecting the output port, the joint spin--resonator density operator obeys the Lindblad master equation ($\hbar=1$)
\begin{equation}
\dot\rho=-i[H,\rho]+\gamma\sum_{i=1}^N\mathcal D[\sigma_i^-]\rho
+\gamma_\phi\sum_{i=1}^N\mathcal D[\sigma_i^z]\rho,
\label{eq:master}
\end{equation}
where $\mathcal D[L]\rho=L\rho L^\dagger-\{L^\dagger L,\rho\}/2$. Local spin reset $\sigma_i^-=(\sigma_i^x-i\sigma_i^y)/2$ occurs at rate $\gamma$, while local pure dephasing in the $z$ basis acts at rate $\gamma_\phi$. The total Hamiltonian is $H=H_E+H_R+H_{ER}$, given by
\begin{align}
H_E&=\frac{h_z}{2}\sum_{i=1}^{N}\sigma_i^z-\frac{J}{N}\sum_{i<j}\sigma_i^x\sigma_j^x,\label{eq:H_E}\\
H_R&=\omega_c a^\dagger a,\qquad H_{ER}=\frac{g}{\sqrt N}(a+a^\dagger)S_z,\nonumber
\end{align}
where $S_z=\frac12\sum_i\sigma_i^z$, $J>0$ denotes the ferromagnetic interaction strength, $h_z>0$ is a static longitudinal field, $\omega_c$ is the resonator frequency, and $g\ge0$ is the collective spin--resonator coupling. The $1/\sqrt N$ scaling ensures an extensive interaction energy when $\langle a\rangle=O(\sqrt N)$. Such collective spin--boson descriptions are widely employed in cavity QED and interacting-spin platforms~\cite{Dicke1954,GrossHaroche1982,Saffman2010,Monroe2021,Schafe2020}, where the nonlinear response of driven open ensembles can support competing stationary and oscillatory attractors~\cite{Mattes2023Bistability,Paulino2026Coupled}.

We adopt homogeneous mean-field dynamics in the thermodynamic limit, factorizing both inter-spin and spin--resonator correlations~\cite{CarolloLesanovsky2021MF,Boneberg2022}. Defining $m_\mu=N^{-1}\sum_i\langle\sigma_i^\mu\rangle$ and $\alpha=\langle a\rangle/\sqrt N=u+iv$, the longitudinal magnetization $m_z$ drives the resonator, while $u$ and $v$ represent its normalized flux and voltage quadratures, respectively. With transverse decoherence rate $\Gamma_2=\gamma/2+2\gamma_\phi$ and passive output damping $\kappa$ included in the voltage quadrature, the closed autonomous equations of motion read
\begin{subequations}\label{eq:mf}
\begin{align}
\dot m_x&=-(h_z+2gu)m_y-\Gamma_2m_x,\\
\dot m_y&=(h_z+2gu+2Jm_z)m_x-\Gamma_2m_y,\\
\dot m_z&=-2Jm_xm_y-\gamma(m_z+1),\\
\dot u&=\omega_cv,\\
\dot v&=-\omega_cu-\tfrac g2m_z-\kappa v.
\end{align}
\end{subequations}
Equivalently, the resonator displacement satisfies the driven damped oscillator equation
\begin{equation}
\ddot u+\kappa\dot u+\omega_c^2u=-\frac{g\omega_c}{2}m_z(t).
\label{eq:osc}
\end{equation}
Neither the Hamiltonian parameters nor the dissipation rates are modulated in time. This homogeneous factorization is exact for the all-to-all model in the thermodynamic limit~\cite{Mattes2025,Benatti2018}. Finite-size effects on the autonomous operating lifetime are analyzed in Sec.~\ref{subsec:output}.

\subsection{Environmental input, output, and cycle work}
\label{subsec:energy_balance}

Pure $z$-dephasing leaves longitudinal spin populations unchanged while suppressing transverse ferromagnetic correlations. Because the ferromagnetic interaction energy is negative, eroding these correlations increases the energy of the working medium. At the mean-field level, $e_J=-Jm_x^2/2$ and $\dot m_x|_\phi=-2\gamma_\phi m_x$, yielding the nonnegative dephasing input power per spin
\begin{equation}
 p_\phi=2J\gamma_\phi m_x^2.\label{eq:dephasing_input}
\end{equation}
A passive load with conductance $G$ extracts coherent electrical power $G\langle\hat V\rangle^2$, which in our dimensionless normalization corresponds to an output power per spin
\begin{equation}
 p_{\mathrm{out}}=2\kappa\omega_cv^2\ge0.
 \label{eq:power}
\end{equation}
The mapping to laboratory units is given by $G/C=\Gamma_{2,\mathrm{ref}}\kappa$, $\langle\hat V\rangle=\sqrt{2N\hbar\Gamma_{2,\mathrm{ref}}\omega_c/C}\,v$, and $P_{\mathrm{out}}=N\hbar\Gamma_{2,\mathrm{ref}}^2p_{\mathrm{out}}$, where $C$ is the resonator capacitance and $\Gamma_{2,\mathrm{ref}}$ is the reference decoherence rate fixed below. Here we account strictly for this coherent output-port power rather than all microscopic thermal fluctuations dissipated inside a resistor~\cite{Monsel2020,Pascal2011}. Distinguishing raw energy transfer from extractable work is foundational to quantum thermodynamics~\cite{Allahverdyan2004,Binder2015} and collective quantum batteries~\cite{AlickiFannes2013,Campaioli2017,Ferraro2018,Rossini2020,Andolina2019}, complementing extensive studies of energy storage and charging across diverse battery architectures~\cite{Farina2019,QuachMunro2020,Quach2022,Hu2022,Joshi2022,Quach2023,Camposeo2025}. In our autonomous engine, sustained performance is quantified by continuous power delivered to the load rather than transient energy stored in the receiver.

The total energy per spin of the loaded mean-field composite is
\begin{equation}
 e_{\mathrm{tot}}=\frac{h_zm_z}{2}-\frac{Jm_x^2}{2}
 +\omega_c(u^2+v^2)+gum_z.
 \label{eq:energy}
\end{equation}
Differentiating Eq.~\eqref{eq:energy} along the trajectories of Eq.~\eqref{eq:mf} yields the exact energy balance
\begin{align}
\dot e_{\mathrm{tot}}&=p_\phi-p_{\mathrm r}-p_{\mathrm{out}},\label{eq:balance}\\
p_{\mathrm r}&=\gamma\left[\left(\frac{h_z}{2}+gu\right)(m_z+1)-\frac{J}{2}m_x^2\right].\label{eq:reset}
\end{align}
The reset energy current $p_{\mathrm r}$ is defined as positive when energy flows out of the composite, though its instantaneous sign can vary over a cycle. The spin-to-resonator energy current is $p_{E\to R}=-g\omega_cvm_z$, satisfying $\dot e_R=p_{E\to R}-p_{\mathrm{out}}$ with resonator energy $e_R=\omega_c(u^2+v^2)$.

Over a closed joint orbit of period $T$, both $e_R$ and $e_{\mathrm{tot}}$ return to their initial values. Defining the integrated energy per cycle $Q_\nu=\int_0^T p_\nu(t)\,dt$, the net work extracted per cycle obeys
\begin{align}
W_{\mathrm{out}}&=Q_\phi-Q_{\mathrm r}\nonumber\\
&=-g\oint m_z\,du=-\frac12\oint m_z\,db,\label{eq:cyclework}\\
b(t)&=h_z+2gu(t).\nonumber
\end{align}
The final equality gives $W_{\mathrm{out}}$ as half the signed area of the $(b,m_z)$ work loop. Whereas externally engineered couplings and charging protocols can shape energy transfer in driven systems~\cite{Ahmadi2024,Centrone2023,Rodriguez2023,Rodriguez2024,Santos2019}, here the effective longitudinal field $b(t)$ is generated internally by the coupled spin--resonator dynamics, meaning the work protocol must emerge self-consistently. Consequently, a stationary fixed point can continuously absorb finite dephasing power while maintaining $v=0$ and $p_{\mathrm{out}}=0$: an energy source alone is insufficient to establish a work-producing cycle.

\section{Self-organized cyclic operation under load}
\label{sec:load_stabilization}
\subsection{Periodic operation retained and restored}
\label{subsec:loaded_states}

The time-crystalline operating state examined here belongs to the broader class of spontaneously organized temporal order~\cite{Wilczek2012,SachaZakrzewski2018,HannafordSacha2022,Zaletel2023,Medenjak2020,ChenZhang2023CTC,CarraroHaddad2024CTC,Huang2025SpinGases,Greilich2026Sync}, where the oscillation period is selected internally by the autonomous dynamics. This contrasts with discrete and prethermal time crystals that lock to an external driving period~\cite{Sacha2015,Else2016,Khemani2016,Pizzi2021,Ye2021,PizziB2021}, as realized across spin and atom--cavity platforms~\cite{Choi2017,Zhang2017,Pal2018,Rovny2018,Kyprianidis2021,Randall2021,OSullivan2020,Kessler2021,Zhu2019,Kongkhambut2024Transition}. Finite-size decay, collective fluctuations, and critical dynamics of dissipative oscillations have also been investigated in open many-body systems~\cite{Owen2018,Carollo2022,Dutta2025,Daviet2025,Sieberer2025}, while related continuous time-crystal platforms have demonstrated oscillatory motion across several physical settings~\cite{Li2024,Liu2023,Wadenpfuhl2023}.

We fix $J=h_z=31.6$, $\gamma=0.3$, and $\gamma_\phi=0.425$, which sets $\Gamma_2=\Gamma_{2,\mathrm{ref}}=1$; time is measured in units of $\Gamma_{2,\mathrm{ref}}^{-1}$ and power per spin in units of $\hbar\Gamma_{2,\mathrm{ref}}^2$. The uncoupled spin limit cycle has fundamental frequency $\Omega_0\simeq6.55$, and we tune the resonator to $\omega_c=2\Omega_0\simeq13.107$ to match the dominant harmonic of the longitudinal magnetization. For each point in the $(g,\kappa)$ parameter map, trajectories are initialized from the \emph{same} spin state and resonator vacuum, $\bm X(0)=(m_x,m_y,m_z,u,v)\simeq(0,-0.0221,-0.627,0,0)$. These maps show the operating states accessible from this fixed preparation.

\begin{figure}[!htbp]
\centering\includegraphics[width=\columnwidth]{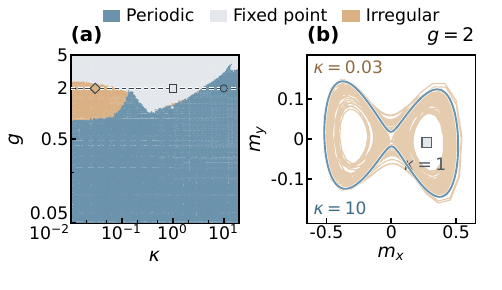}
\caption{Load-controlled operating states.
(a) Operating states reached from a common initial condition. Sand denotes irregular or unresolved trajectories.
(b) Phase-space projections onto the $m_x$--$m_y$ plane at the three marked points on the $g=2$ line in (a). The irregular trajectory is shown over $300\le t\le400$.}
\label{fig:load_stabilization}
\end{figure}

Figure~\ref{fig:load_stabilization}(a) illustrates two distinct pathways to periodic operation under load. At sufficiently weak coupling, the collective spin oscillation persists after the output port is connected across a broad range of load damping. At stronger coupling, resonator backaction can destabilize this cycle into a fixed point or an irregular trajectory; however, increasing the load damping restores accessible periodic motion. This load-induced restoration constitutes the central collective feedback effect. The three representative points along $g=2$ in Fig.~\ref{fig:load_stabilization}(b) highlight this contrast at fixed coupling: $\kappa=0.03$ yields an irregular long-time orbit, $\kappa=1$ collapses onto a stationary fixed point, and $\kappa=10$ settles into a closed, stable limit cycle. The stabilized orbit at $\kappa=10$ has period $T\simeq0.993$ and maximum nontrivial Floquet multiplier modulus $r_F\simeq0.281$, obtained by shooting the full five-dimensional orbit and integrating its variational equations.

The restoration of periodic operation can be understood as a suppression of excessive resonator feedback by the load. For a magnetization component $m_{z,\nu}e^{i\nu t}$, the oscillatory feedback field $\delta b_\nu=2g u_\nu$ has gain
\begin{equation}
 \mathcal G(\nu)=\frac{g^2\omega_c}
 {\left|\omega_c^2-\nu^2+i\kappa\nu\right|}.
 \label{eq:susceptibility}
\end{equation}
Near resonance, when damping dominates the detuning, this gain scales as $g^2/\kappa$. At stronger spin--resonator coupling, a weakly damped resonator can exert strong backaction that disrupts the intrinsic dephasing-sustained cycle. Increasing the load damping reduces this feedback gain and, in the regimes illustrated in Fig.~\ref{fig:load_stabilization}, allows stable periodic operation to recover. The load therefore acts both as an energy sink and as a regulator of the resonator feedback governing autonomous cyclic operation. Appendix~\ref{app:load_feedback} derives the corresponding feedback response and its contribution to the collective dynamics. The reference fixed point remains linearly stable at both $\kappa=1$ and $\kappa=10$, demonstrating coexistence of the restored cycle and stationary operation~\cite{YangZhang2026}.

For numerical classification across the parameter plane, we use adaptive high-accuracy DOP853 integration, monitoring full-state recurrence, amplitude and period drift, and convergence of integrated energy currents over extended time windows up to $t=6400$ where required. Representative stable cycles are independently verified via boundary-value shooting and their nontrivial Floquet multipliers, $r_F=\max_{\mu_j\ne1}|\mu_j|$.

\subsection{From collective motion to output power}
\label{subsec:output}

\begin{figure}[!htbp]
\centering\includegraphics[width=\columnwidth]{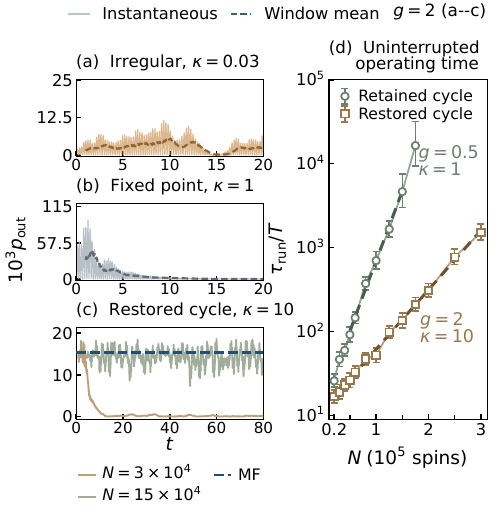}
\caption{Power output and finite-size operating endurance.
(a,b) Instantaneous and window-averaged mean-field output.
(c) One-period moving averages for two finite-$N$ stochastic trajectories starting from the same loaded-cycle state, shown alongside the mean-field cycle average.
(d) Uninterrupted running time $\tau_{\rm run}/T$ versus system size $N$ for the retained $(g,\kappa)=(0.5,1)$ and restored $(2,10)$ cycles, where $T$ is the corresponding mean-field period.
Here $\tau_{\rm run}$ is the mean running time to a sustained drop in the window-averaged output power below $25\%$ of the mean-field cycle average.
The means are estimated from independent trajectories with right censoring at the observation cutoff.
Error bars show 95\% confidence intervals. Dashed lines show exponential fits over the indicated ranges. }
\label{fig:instantaneous_output}
\end{figure}

Macroscopic collective oscillations benefit the engine only to the extent that they couple energy into the output port. To separate the rapid LC carrier oscillations from their slower envelope, we define a trailing-window average of duration $\Delta$ for the instantaneous powers in Eqs.~\eqref{eq:dephasing_input} and~\eqref{eq:power}:
\begin{equation}
 p_\nu^{(\Delta)}(t)=\frac1\Delta\int_{t-\Delta}^{t}p_\nu(s)\,ds,
 \qquad \nu=\phi,\mathrm{out}.
\label{eq:active_power}
\end{equation}
When $\Delta$ matches an integer multiple of the period $T$, this quantity reduces to the constant cycle-averaged power of a stable orbit. For uniform comparison in Fig.~\ref{fig:instantaneous_output}, we set $\Delta=T_{\mathrm{ref}}\simeq0.993$ and evaluate the average directly from the cumulative energy counter.

This dynamical contrast has direct consequences for the energy budget. At the stationary point $(g,\kappa)=(2,1)$, the spins continuously absorb dephasing power, $p_\phi\simeq2.0124$, yet $p_{\mathrm{out}}=0$ and $p_{\mathrm r}=p_\phi$: the environment injects energy without producing sustained macroscopic electrical work. At $(g,\kappa)=(2,10)$, by contrast, the stabilized cycle converts part of its dephasing input into continuous output power, $\overline p_{\mathrm{out}}\simeq0.0153$. The loaded limit cycle thus operates as a self-organized work protocol enclosing a finite work-loop area in Eq.~\eqref{eq:cyclework}. Irregular trajectories can also yield finite mean power. The periodic attractor organizes this output into a reproducible cycle with well-defined work per cycle.

To characterize the endurance of work production at finite $N$, we follow the semiclassical spin-diffusion description of Ref.~\cite{YangZhang2026}, retaining the dynamical response of the loaded LC mode:
\begin{subequations}\label{eq:finiteN}
\begin{align}
d\bm m&=\bm F(\bm m,u)\,dt+N^{-1/2}B(\bm m)\,d\bm W_t,\\
BB^{\mathsf T}&=\begin{pmatrix}
2\Gamma_2&0&\gamma m_x\\
0&2\Gamma_2&\gamma m_y\\
\gamma m_x&\gamma m_y&2\gamma(1+m_z)
\end{pmatrix}.
\end{align}
\end{subequations}
Here local spin fluctuations feed back into the otherwise deterministic resonator dynamics. We define $\tau_{\rm run}$ as the time to the first sustained interruption of cyclic power delivery. Figure~\ref{fig:instantaneous_output}(c) contrasts a smaller ensemble whose cyclic output is interrupted with a larger one that continues running over the same observation window. Figure~\ref{fig:instantaneous_output}(d) maps the size dependence of $\tau_{\rm run}$ for both the weak-coupling retained cycle and the load-restored cycle. The characteristic running time increases rapidly with $N$ and is described over the fitted ranges by $\tau_{\rm run}\propto\exp(N/N_*)$, with fitted $N_*\simeq2.6\times10^4$ and $6.1\times10^4$, respectively. The weak-coupling cycle generates less mean power but lasts longer before a sustained interruption over the common size range. Although phase diffusion limits the coherence time, the engine continues to deliver cyclic work unless its trajectory is captured by a fixed point.

\section{Energy extraction and dynamical reliability}
\label{sec:efficiency}

Over a common averaging window $\Delta$, we quantify the fraction of dephasing input delivered to the output port via the energy-extraction efficiency
\begin{equation}
\eta_\phi^{(\Delta)}(t)=\frac{\int_{t-\Delta}^t p_{\mathrm{out}}(s)\,ds}
{\int_{t-\Delta}^t p_\phi(s)\,ds},\qquad
\int_{t-\Delta}^t p_\phi\,ds>0.
\label{eq:efficiency}
\end{equation}
Its asymptotic value $\eta_\phi$ is evaluated from converged integrated energy currents or over one complete period of a verified limit cycle. $\eta_\phi$ quantifies the \emph{dephasing-to-output energy-extraction efficiency}, with the costs of maintaining the reservoirs and control resources excluded from its definition~\cite{Weber2026,Chiribella2021}. The operating-state classification distinguishes periodic from irregular power delivery. At finite size, the operating lifetime governs the duration of cyclic extraction, while phase diffusion can preserve nonzero mean power.

\begin{figure}[!htbp]
\centering\includegraphics[width=\columnwidth]{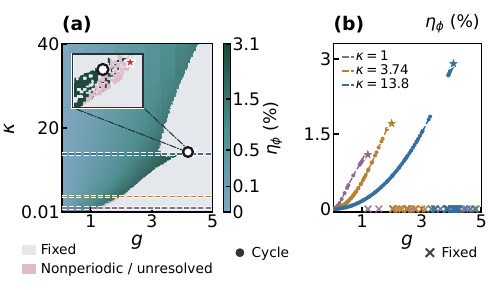}
\caption{Energy-extraction efficiency and operating states.
(a) Blue--teal shading shows $\eta_\phi$ for points classified as periodic.
The red star marks the highest sampled long-time efficiency, attained at a nonperiodic point.
The black-edged circle marks a stable limit cycle.
(b) Efficiency versus $g$ at fixed $\kappa$.
Dashed lines connect adjacent samples classified as periodic, and stars mark the most efficient periodic sample on each cut.}
\label{fig:efficiency}
\end{figure}

Figure~\ref{fig:efficiency} maps $\eta_\phi$ across the $(g,\kappa)$ plane and compares fixed-$\kappa$ cuts while explicitly distinguishing dynamical operating states. Blue--teal shading indicates points classified as periodic via full five-dimensional state recurrence within converged energy windows; gray marks fixed points, and rose denotes irregular or unresolved trajectories, including high-output chaotic states exhibiting positive numerical Lyapunov exponents. Finite-time recurrence identifies the colored operating regions; Floquet analysis verifies representative stable cycles. The inset resolves the narrow high-output region.

For a given load damping $\kappa$, the spin--resonator coupling $g$ can be tuned to optimize energy-extraction efficiency while retaining periodic operation. Along each fixed-$\kappa$ cut in Fig.~\ref{fig:efficiency}(b), increasing $g$ initially improves energy transfer to the resonator and raises $\eta_\phi$. At stronger coupling, resonator backaction can change the operating state reached from the same preparation. The stars mark the highest efficiency among the sampled periodic states for each load.

Tracking the high-efficiency period-one branch continuously from $(g,\kappa)\simeq(4.19,14.269)$ toward $(4.195,14.274)$ reveals that the leading nontrivial Floquet multiplier crosses $-1$ near $g\simeq4.195$, signaling a period-doubling instability. After the original cycle loses stability, higher-period or irregular trajectories can still deliver output power, echoing broader studies distinguishing regular synchronization from complex nonlinear temporal order~\cite{Solanki2024,Nadolny2025,Greilich2025}.

Across our sampled grid, the largest converged \emph{mean} energy-extraction efficiency reaches approximately $3.093\%$ at $(g,\kappa)=(4.244,14.32)$, but long-time variational analysis yields a positive largest Lyapunov exponent of roughly $0.6$ (in units of $\Gamma_{2,\mathrm{ref}}$), ruling out a stable simple cycle. By contrast, an independently closed periodic orbit near $(g,\kappa)=(4.19,14.269)$ achieves $\eta_\phi\simeq3.02\%$ with period $T\simeq4.39$ and Floquet modulus $r_F\simeq0.52<1$. This verified high-efficiency cyclic state provides a stable operating reference against the larger nonperiodic mean-efficiency peak. Combining the efficiency and operating-state maps identifies suitable couplings for cyclic work at a prescribed load.

\section{Conclusion}
\label{sec:conclusion}

Within homogeneous mean-field dynamics, we have established a dephasing-powered autonomous spin engine whose collective motion has generated time-crystalline work cycles. We have identified two routes to cyclic operation: preservation of the unloaded spin limit cycle under weak resonator backaction and restoration of a stable joint spin--resonator cycle through load damping at stronger coupling. We have traced the latter mechanism to load-dependent suppression of resonator feedback, and we have connected both operating regimes to their delivered power and dephasing-to-output efficiency.

Using semiclassical finite-size simulations, we have quantified the rapid growth of uninterrupted operating times with spin number and have resolved distinct lifetime dependences for the retained and restored cycles. Together, these results have demonstrated how many-body interactions, dephasing, and load feedback have enabled the self-organization of stationary environmental input into sustained cyclic work. Further studies can examine the resulting engine performance in other interacting open-system platforms.

\begin{acknowledgments}
This work was supported by the National Natural Science Foundation of China (Grants No. 12375013 and No. 12547109), the Guangdong Basic and Applied Basic Research Fund (Grant No. 2023A1515011460), and Guangdong Provincial Quantum Science Strategic Initiative (Grant No. GDZX2503008).
\end{acknowledgments}

\appendix
\section{Load regulation of collective feedback}
\label{app:load_feedback}

The load regulates collective feedback by changing the resonator response that returns the longitudinal magnetization to the spins as a field. Writing $\delta b(t)=2gu(t)$ and $b(t)=h_z+\delta b(t)$, Eq.~\eqref{eq:osc} gives
\begin{equation}
\delta\ddot b+\kappa\,\delta\dot b+\omega_c^2\delta b
=-g^2\omega_c m_z(t).
\label{eq:feedback_field_ode}
\end{equation}
Thus $m_z$ drives the resonator, whose displacement acts back on the spins. The damping rate $\kappa$ enters this closed feedback channel directly.

For $\kappa>0$, after the homogeneous resonator transient has decayed, the returning field can be written as
\begin{equation}
\begin{aligned}
\delta b(t)&=-g^2\omega_c\int_0^\infty
G_\kappa(\tau)m_z(t-\tau)\,d\tau,\\
\widetilde G_\kappa(s)&=\frac{1}{s^2+\kappa s+\omega_c^2}.
\end{aligned}
\label{eq:feedback_memory}
\end{equation}
Here $G_\kappa$ is the causal Green function of the damped resonator, and $\widetilde G_\kappa$ is its Laplace transform. The returning field therefore depends on the history of the longitudinal magnetization, with a memory kernel controlled by the load.

Using the Fourier convention $e^{i\nu t}$, Eq.~\eqref{eq:feedback_field_ode} yields
\begin{equation}
\delta b_\nu=K_\kappa(\nu)m_{z,\nu},\qquad
K_\kappa(\nu)=-\frac{g^2\omega_c}{\Delta_\nu+i\kappa\nu},
\label{eq:feedback_transfer}
\end{equation}
where $\Delta_\nu=\omega_c^2-\nu^2$. For $\nu>0$, the gain and the phase relative to $m_{z,\nu}$ are
\begin{equation}
\begin{aligned}
|K_\kappa(\nu)|&=\frac{g^2\omega_c}
{\sqrt{\Delta_\nu^2+\kappa^2\nu^2}},\\
\theta_\kappa(\nu)&=\pi-\arg(\Delta_\nu+i\kappa\nu).
\end{aligned}
\label{eq:feedback_gain_phase}
\end{equation}
Increasing $\kappa$ reduces the transfer gain at a fixed nonzero frequency and moves the phase of detuned components toward a quadrature response. When $|\Delta_\nu|\ll\kappa\nu$, the gain is approximately $g^2\omega_c/(\kappa\nu)$; near $\nu=\omega_c$, it reduces to $g^2/\kappa$. The magnetization amplitudes and frequencies themselves remain selected by the joint dynamics. The zero-frequency response gives $\bar b=h_z-g^2\bar m_z/\omega_c$. Although this relation contains no explicit $\kappa$, the static shift depends indirectly on the load through the self-consistent $\bar m_z$.

To make the action of this returning field on the intrinsic collective feedback explicit, let $x=m_x$ and define the shifted energy variable
\begin{equation}
Q_b=b(t)(m_z+1)-Jx^2.
\label{eq:loaded_energy_variable}
\end{equation}
The spin equations in Eq.~\eqref{eq:mf} imply
\begin{equation}
\begin{aligned}
\dot Q_b&=-\gamma Q_b+4J\gamma_\phi x^2
+(m_z+1)\dot b,\\
\dot b&=2g\omega_cv.
\end{aligned}
\label{eq:loaded_energy_feedback}
\end{equation}
The first two terms retain the energy input and relaxation structure of the unloaded model. The field-rate term adds the resonator response to the evolution of the internal feedback variable.

On trajectories for which $b(t)\ne0$, the relations $m_y=-(\dot x+\Gamma_2x)/b$ and $m_z=(Q_b+Jx^2)/b-1$ eliminate $m_y$ and $m_z$ from the transverse equations. Defining $r(t)=\dot b/b$, one obtains
\begin{equation}
\begin{aligned}
\ddot x+(2\Gamma_2-r)\dot x
&+\bigl(b^2+\Gamma_2^2-2Jb-\Gamma_2r\bigr)x\\
&+2J^2x^3+2JQ_bx=0.
\end{aligned}
\label{eq:loaded_reduced_oscillator}
\end{equation}
At $g=0$, $b=h_z$ is constant, and Eqs.~\eqref{eq:loaded_energy_feedback} and~\eqref{eq:loaded_reduced_oscillator} recover the unloaded feedback loop: $x^2$ drives the energy variable, which acts back on the transverse oscillation through $-2JQ_bx$. Under loading, the amplitude and phase of the returning field modify $b(t)$, while $r(t)$ enters the derivative and restoring-force terms. These time-dependent coefficients and the evolution of $Q_b$ jointly determine the loaded orbit.

The corresponding cycle balance follows by multiplying Eq.~\eqref{eq:loaded_reduced_oscillator} by $\dot x$ and integrating along a closed joint orbit of period $T$:
\begin{equation}
\begin{aligned}
2\Gamma_2\int_0^T\dot x^2\,dt
={}&-J\oint Q_b\,d(x^2)
+\int_0^T r\dot x^2\,dt\\
&-\frac12\oint
\bigl(b^2-2Jb-\Gamma_2r\bigr)\,d(x^2).
\end{aligned}
\label{eq:loaded_feedback_cycle_balance}
\end{equation}
The first term is the contribution of the internal energy variable; the remaining terms arise from the time-dependent longitudinal field. At $g=0$, the latter vanish and the unloaded feedback balance is recovered. Equation~\eqref{eq:loaded_feedback_cycle_balance} describes the damping and feedback balance of the reduced transverse motion.

The physical joint energy is $e_{\mathrm{tot}}=(Q_b-b)/2+\omega_c(u^2+v^2)$. Its balance remains Eq.~\eqref{eq:balance}, with $p_\phi=2J\gamma_\phi x^2$, $p_{\mathrm r}=\gamma Q_b/2$, and $p_{\mathrm{out}}=2\kappa\omega_cv^2$. The reset current is defined as positive out of the composite. Averaging over a closed joint cycle gives
\begin{equation}
2J\gamma_\phi\,\overline{x^2}
=\frac{\gamma}{2}\,\overline{Q_b}
+\overline p_{\mathrm{out}},\qquad
\overline f=\frac1T\int_0^T f(t)\,dt.
\label{eq:loaded_physical_cycle_balance}
\end{equation}

\end{document}